\documentclass[conference]{IEEEtran}

\IEEEoverridecommandlockouts
\usepackage{acronym}
\usepackage[caption=false]{subfig}
\usepackage{booktabs}
\usepackage{multirow}
\usepackage{cite}
\usepackage{amsmath,amssymb,amsfonts}
\usepackage{graphicx}
\usepackage{textcomp}
\usepackage{xcolor}
\usepackage{balance}
\usepackage[capitalise]{cleveref}
\def\BibTeX{{\rm B\kern-.05em{\sc i\kern-.025em b}\kern-.08em
    T\kern-.1667em\lower.7ex\hbox{E}\kern-.125emX}}
\newcommand{\mtx}[1]{\mathbf{#1}}

\def\A{{\mtx A}}

\def\s{{\mtx x}}
\def\x{{\mtx x}}
\def\n{{\mtx n}}
\def\p{{\mtx p}}

\acrodef{ML}{Machine Learning}
\acrodef{KNN}{K Nearest Neighbors}
\acrodef{RF}{Random Forest}
\acrodef{CPPS}{Cyber-Physical Production Systems}
\acrodef{GNSS}{Global Navigation Satellite System}
\acrodef{RSSI}{Received Signal Strength Indicator}
\acrodef{SVC}{Support Vector Classifier}
\acrodef{SNR}{Signal-to-Noise Ratio}
\acrodef{SVM}{Support Vector Machine}
\acrodef{SAL-RB-Dataset}{SAL Autarkic Localization RSSI BLE Dataset}
\acrodef{EPhESOS}{Energy and Power Efficient Synchronous Sensor Network} 
\acrodef{TDMA}{Time Division Multiple Access}
\acrodef{BLE}{Bluetooth %\textsuperscript{\textregistered} 
Low Energy}

\acrodef{UWB}{Ultra Wideband}
\title{RSSI-Based Location Classification Using \\a Particle Filter to Fuse Sensor Estimates
\thanks{This work is funded by the InSecTT project (https://www.insectt.eu/). InSecTT has received funding from the ECSEL Joint Undertaking (JU) under grant agreement No 876038. The JU receives support from the European Union’s Horizon 2020 research and innovation programme and Austria, Sweden, Spain, Italy, France, Portugal, Ireland, Finland, Slovenia, Poland, Netherlands, Turkey. The document reflects only the author’s view and the Commission is not responsible for any use that may be made of the information it contains.}
}
\author{\IEEEauthorblockN{Thomas Blazek$^\star$, Julian Karoliny$^\star$, Fjolla Ademaj$^\star$, Hans-Peter Bernhard$^{\star,\dagger}$}
\IEEEauthorblockA{$^\star$ Silicon Austria Labs Gmbh,
Research Unit Wireless Communications,\\
$^\dagger$ Johannes Kepler University Linz, Institute for Communications Engineering and RF-Systems,\\
Altenberger Straße 69, Linz, Austria\\
}}

\begin{document}
%\ninept
%
\maketitle
\begin{abstract}
  For Cyber-Physical Production Systems (CPPS), localization is becoming increasingly important as wireless and mobile devices are considered an integral part. While localizing targets in a wireless communication system based on the Received Signal Strength Indicator (RSSI) of transmitted beacons is a well-known strategy, it is often limited by the quality of the RSSI sensors We propose to use a particle filter that fuses RSSI measurements of different sensors. This allows us to incorporate sensor non-idealities in our model, and achieve a high-quality position estimate that is not limited by them. The estimation performance is evaluated using real-world measurements of a car in a chamber.
  In the second step, we  use Machine Learning (ML) to classify where the vehicle is. Our results show that the location output of the particle filter is a better input to the ML technique than the raw RSSI data, and we achieve improved classification accuracy while simultaneously reducing the number of features that the ML has to consider. We furthermore compare the performance of multiple ML algorithms and demonstrate that SVMs provide the overall best performance for the given task.
  %
  %For Cyber Physical Production System (CPPS), localization is becoming increasingly important as wireless and mobile devices are considered an integral part. While localizing targets in a wireless system based on the Received Signal Strength Indicator (RSSI) of transmitted beacons is a well-known strategy, it is often limited by the quality of the RSSI sensors. We propose to use a particle filter that fuses RSSI values of different sensors to incorporate sensor non-idealities in our model and achieve a high-quality position estimate. The estimation performance is evaluated using real-world data of an automotive testbed.
  % We use Machine Learning (ML) with particle filter location input to classify the vehicle position which performs out ML with raw RSSI input data. Furthermore, we improved classification accuracy while simultaneously reducing feature number. A performance comparison of multiple ML algorithms demonstrates that SVMs provide the overall best performance for the given task.
  %
\end{abstract}
\section{Introduction}
\label{sec:intro}
In \ac{CPPS}, knowing the location of moving objects, be it robots or vehicles, is already often essential, and with the advent of cyber-physical systems and smart agents, this issue is expected to become even more pressing~\cite{leitaoSmartAgentsIndustrial2016}. Therefore, a significant effort is being put into the ability to localize objects in such a setting.
Broadly speaking, there are systems specifically designed for this task, and systems where the
localization is done as a side benefit of another system. Dedicated systems include radar \cite{ebeltCooperativeIndoorLocalization2014}, as well as \ac{GNSS}  systems. While the former require a distributed net of radar sensors, the latter pose large challenges when deployed indoor \cite{nirjonCOINGPSIndoorLocalization2014,yassinRecentAdvancesIndoor2017}.
% Furthermore, \ac{UWB} systems allow accurate localization \cite{wangPrototypingExperimentalComparison2015}, but are rarely already deployed in such a setting. 

Alternatively, existing systems are exploited. There, the most obvious choices are visual
tracking through the use of camera systems, and exploiting radio communications that are deployed
in the tracked object.
Using video systems for tracking is currently of high interest within the machine and deep learning
community \cite{ciaparroneDeepLearningVideo2020,xuDeepLearningMultiple2019}. A major downside of
the visual approach is that the camera system must be adapted to cover the relevant area, and
a traceable object does nothing to initiate being tracked. On the other hand, RF-based localization
works based on beacons transmitted from a traceable object and is therefore harder to miss.
Estimation of location based on the \ac{RSSI} has been studied for some time \cite{wangRSSIBasedBluetoothIndoor2015,mistryRSSIBasedLocalization2015,
  yaghoubiEnergyEfficientRSSIBasedLocalization2014}. The \ac{RSSI} based estimation usually
resorts to variants of triangulation or multilateration~\cite{zafariSurveyIndoorLocalization2019}. However, due to the often limited dynamic
range of RSSI estimates, and the noise floor of receivers, multilateration may not be a solution, and authors rarely apply more advanced schemes such as maximum likelihood estimation
\cite{zancaExperimentalComparisonRSSIbased2008}.

\subsection{Contribution}
We present an approach that allows us to localize a target transmitting a Bluetooth beacon using distributed low-cost sensors with limited dynamic range and high noise floor. Similar to the idea sketched but not implemented in \cite{wuParticleFilterSupport}, we employ a sensor fusion particle filter \cite{reppTargetTrackingUsing2018} to convert a large number of low-quality sensor measurements into one high-quality position and velocity estimate. We also account for the sensor placement on a vehicle resulting in a non-omnidirectional antenna pattern. Our results, which are based on real-world measurements, show that this choice allows accurately fusing the highly imperfect sensor data. This is presented in \cref{sec:particlefiltering}.

We then use the high-quality position estimates to classify three states that are especially relevant to us, by using \ac{ML} techniques. We compare three different classifiers, \ac{SVM}, \ac{KNN} and \ac{RF} on the location estimates. Furthermore, we analyse the influence of data-prescaling. This two-step process improves the classification accuracy while simultaneously reducing the number of required features for the Machine learning model. The results are shown in \cref{sec:svm}.
\subsection{Notation}
Scalars are written as \(x\), while vectors and matrices are denoted as lower- and uppercase boldface respectively (e.g. \(\x\) and \(\mtx X\)). Time indices are indicated using square brackets. The Euclidean norm is written as $\|\cdot\|$. $x[t]$ denotes the sample of $x$ at discrete-time index $t$.
\label{sec:format}
\section{System Model}
\label{sec:system_model}
\begin{figure}
  \centering
  \includegraphics[width=0.5\textwidth]{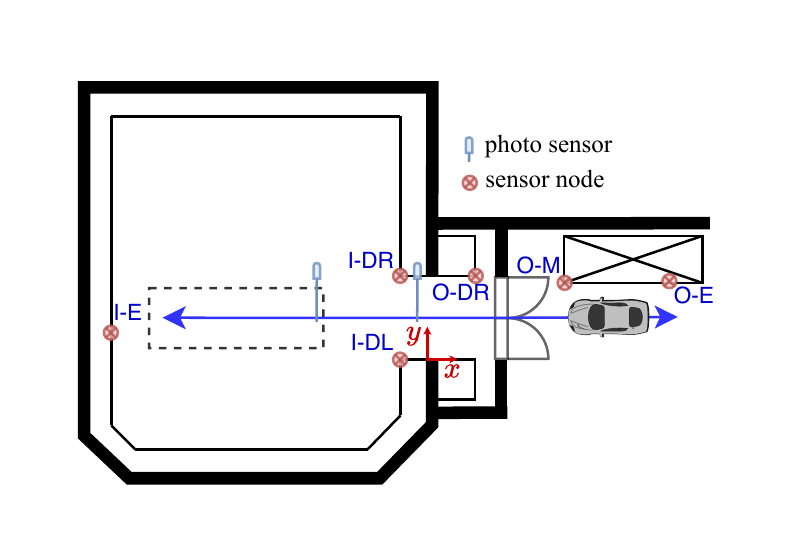}
  \caption{Measurement Setup. When the car is fully to the left of the left photo sensor, we assign state 2. If it is completely right of the right photo sensor, we assign state 0. Otherwise, we assign state~1.}\label{fig:measurement_setup}
\end{figure}
We base our analysis on real-world measurements taken from the \ac{SAL-RB-Dataset} \cite{SAL-RB_Dataset}.
This data set presents the scenario as given in \cref{fig:measurement_setup}.
A car moves in and out of a chamber along the \(x\)-axis, where the positive direction points outwards of the chamber. From the data set's provided photo sensor data, we compute labels for three distinct states: the car being outside of the chamber (right of the rightmost photo sensor) is designated state \(0\), the car being inside the chamber in the end position fully past the left photo sensor is state \(2\), and the transition region where the car has entered the chamber but not yet reached the defined position is designated state \(1\).
The car transmits periodically at an interval of \(\Delta t = 100\)\,ms using the \ac{EPhESOS} protocol and the \ac{BLE} physical layer~\cite{life_cycle_ephesos}. This \ac{TDMA} protocol allows exact time synchronization of the measurement series. At six positions, sensors are placed that record the \ac{RSSI} of the transmitted Bluetooth beacons. These sensors are low-cost in nature, and therefore have a very limited dynamic range and \ac{SNR}.

\begin{figure}[t]
  \centering
  \includegraphics[width=\linewidth]{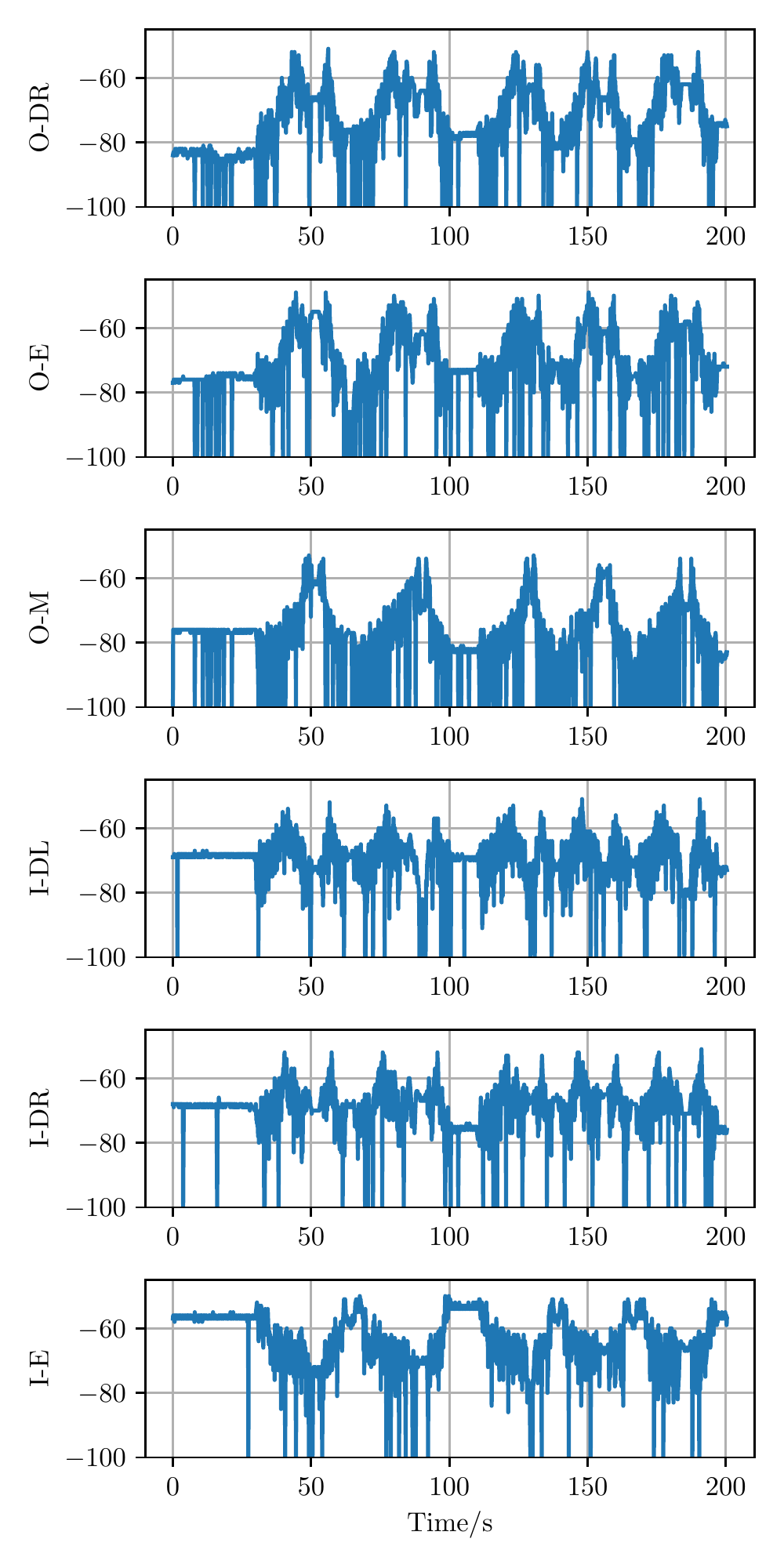}
  \caption{Exemplary RSSI measurements in dBm of the \(6\) used sensors. }\label{fig:sensor_data}
\end{figure}

We use six sets of measurements, that all
reflect the described scenario. Each measurement contains between two and five drives in and out of the chamber. \Cref{fig:sensor_data} shows the \ac{RSSI} data in dBm of one such measurement. We refer to the three-dimensional position and velocity of the car as \(\mtx p\) and \(\mtx v\).
In our setup, the car moves purely along the \(x\)-axis. Hence, we introduce the vector $\s$ describing the cars state, which is comprised of the current position and velocity along the $x$-axis
\begin{equation}
  \s = \begin{bmatrix}
    p^x \\
    v^x\end{bmatrix}\,.\label{eq:state_vector}
\end{equation}
While the car will accelerate and decelerate at the ends of the movement regions, we assume that a uniform movement model holds for the majority of the time
\begin{equation}
  \s[t+1] = \underbrace{\begin{bmatrix}
      1 & 1 \\
      0 & 1
    \end{bmatrix}}_{\A}\s[t]\,.
\end{equation}
Given the positions of the transmitter \(\p\) and the \(s\)\,th sensor \(\mtx q_s\), the signal power at the sensor is provided in dB by the pathloss equation

\begin{align}
  l(\p, \mtx q_s) = & P_{\text{tx}} - {PL}_0 \notag                                                                              \\
                    & - n10\log_{10}\left(\frac{\|\mtx q_s - \p\|}{1\,\text{m}}\right) + \pi(\p, \mtx q_s)\,,\label{eq:pathloss}
\end{align}
where \(\text{PL}_0\) is the pathloss at \(1\)\,m, \(n\) is the pathloss exponent, and \(\pi(\p, \mtx q_s)\) represents the antenna pattern of the transmitter. Since the transmitter is mounted on the front side of a car, we expect significant directionality, even if the mounted antenna is originally omnidirectional. The \ac{RSSI} estimate at the receiving sensor nodes is calculated in dB, and due to the low-cost nature, significantly noisy. Furthermore, we assume that below a threshold \(P_{\text{floor}}\), the sensor does not capture the pathloss anymore, and instead reports a noisy realization of the noise floor.
Hence, we assume that the actual likelihood function for a given \ac{RSSI} measurement \(P_\text{rx}\) in dB is normal distributed
\begin{align}
  f(P_{\text{rx}}|\p, \mtx q_s) = & \frac{1}{\sqrt{2\pi\sigma^2}}\exp\left(-\frac{(P_{\text{rx}} - \mu_P)^2}{2\sigma^2}\right)\,,\notag \\
  \mu_P =                         & \max\big(P_{\text{floor}}, l(\p, \mtx q_s)\big)\,.
\end{align}
As \(p^y\) and \(p^z\) are constant throughout the measurement, only the \(p^x\) coordinate is required, and \( f(P_{\text{rx}}|\p, \mtx q_s) =  f(P_{\text{rx}}|\s, \mtx q_s) \).
\section{Localization via Sensor Fusion Particle Filtering}
\label{sec:particlefiltering}
In this section, we estimate the car position from the low-quality \ac{RSSI} recordings of the sensors. The particle filter, as well as the sensor fusion, is taken from \cite{ashury2020accuracy}. Fundamentally, the filter is initialized once, and then the steps \emph{prediction}, \emph{update}, and \emph{resampling} are cyclically executed.
To \emph{initialize},
we use \(n=300\) particles that have the shape given in \cref{eq:state_vector}.
\paragraph{Initialization} We draw \(n\) \(p^x\) from a uniform distribution \(\mathcal{U}(-25,25)\), and \(v^x\) from \(\mathcal{U}(-3,3)\).
Each particle is associated with a weight \(w_i\) with \(i \in \mathbb{N} \) and \( i \in [1,n] \) that is initially set to \(w_i=n^{-1}\).
\paragraph{Predition} For every particle \(\p_i\), we compute the \emph{prediction}
\begin{equation}
  \s_i[t+1] = \A\s_i[t] + \n[t]\,.
\end{equation}
\(\n[t]\) is a multivariate normal driving noise term that is zero-mean with a covariance matrix
\begin{equation}
  \mtx C = \begin{bmatrix}
    4 & 0 \\
    0 & 4
  \end{bmatrix}\,.
\end{equation}
\paragraph{Update} Now, we \emph{update} the estimates based on the recorded measurements. For every sensor \(s\), an \ac{RSSI} value \(P_s[t]\) is recorded, unless the sensor refuses to provide a measurement at that point. We then calculate the local likelihood function from sensor \(s\) to particle \(p_i\) as the conditional probability density \( f_s^{(i)}(P_s[t]|\s_i, \mtx q_s)\). If a sensor does not provide a measurement, this term will be set to \(1\). Then, the sensor fusion is performed by updating the weights for the particles according to
\begin{align}
  w_i' = & \prod_{s\in \mathcal{S}}f_s^{(i)}(P_s[t]|\s_i, \mtx q_s)\,,\notag \\
  w_i =  & \frac{w_i'}{\sum_jw_j'}\,,
\end{align}
where \(\mathcal{S}\) is the set of all existing sensors.
Now an estimate of the target, as well as the estimation variance is computed as
\begin{align}
  \hat{\s} =           & \sum_iw_i\s_i\,,\notag          \\
  \hat{\mtx\sigma}^2 = & \sum_iw_i(\s_i - \hat{\s})^2\,.
\end{align}
\paragraph{Resampling} To improve the estimation quality, a new set of \(n\) particles is \emph{resampled} from the old set with replacements with the probability of drawing \(\p_i\) being \(w_i\). Afterwards, the weights are reset to \(w_i=n^{-1}\), and the next iteration starts at \emph{prediction}.

%\subsection{Fitting parameters from measurements}
\begin{figure}[t]
  \begin{minipage}[b]{1.0\linewidth}
    \centering
    \centerline{\includegraphics[width=1.0\linewidth]{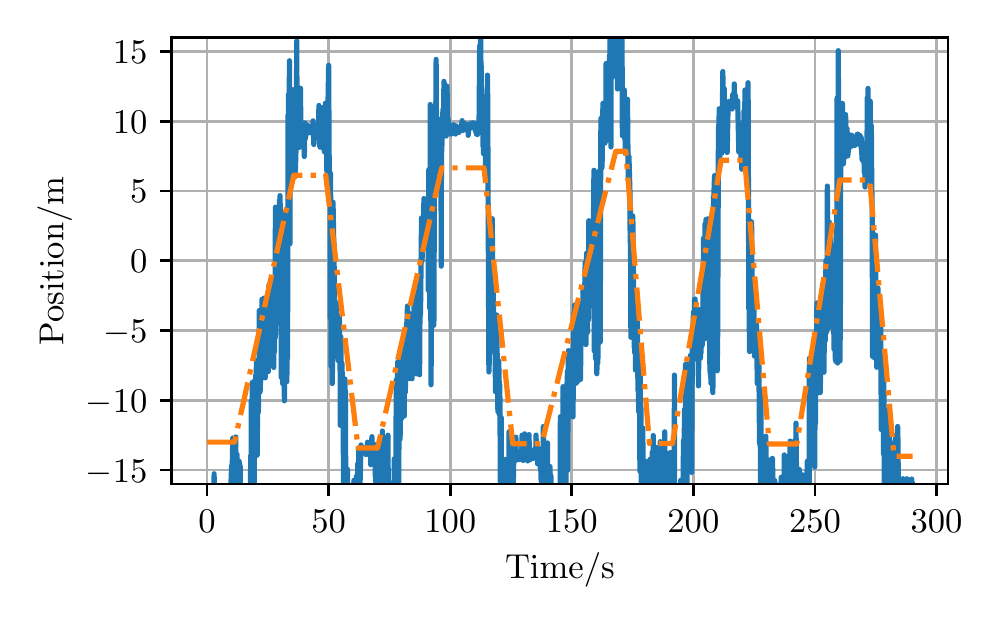}}
    \centerline{(a) Omnidirectional estimation.}\smallskip
  \end{minipage}
  \begin{minipage}[b]{1.0\linewidth}
    \centering
    \centerline{\includegraphics[width=1.0\linewidth]{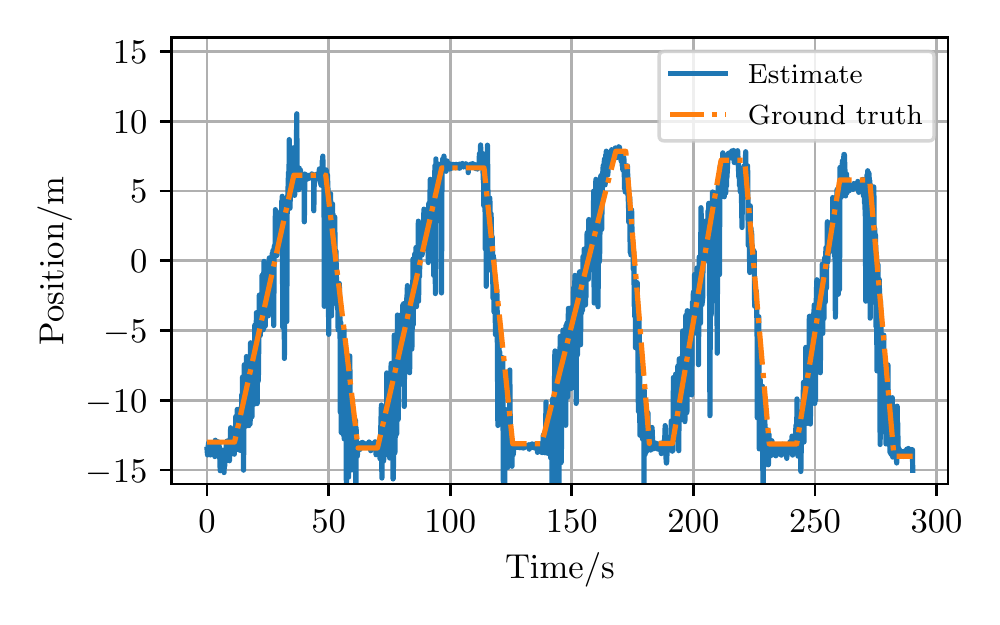}}

    \centerline{(b) Directional estimation using \cref{eq:directional}.\label{fig:pf_b}}
  \end{minipage}

  \caption{Performance of the position estimation compared to the ground truth. (a) shows the filter with an omnidirectional pattern, while (b) uses the antenna model from \cref{eq:directional}.\label{fig:pf_results}}
\end{figure}

\begin{figure*}[t]
  \centering
  \subfloat[SVM, Standard Scaler]{\includegraphics[width=.5\linewidth]{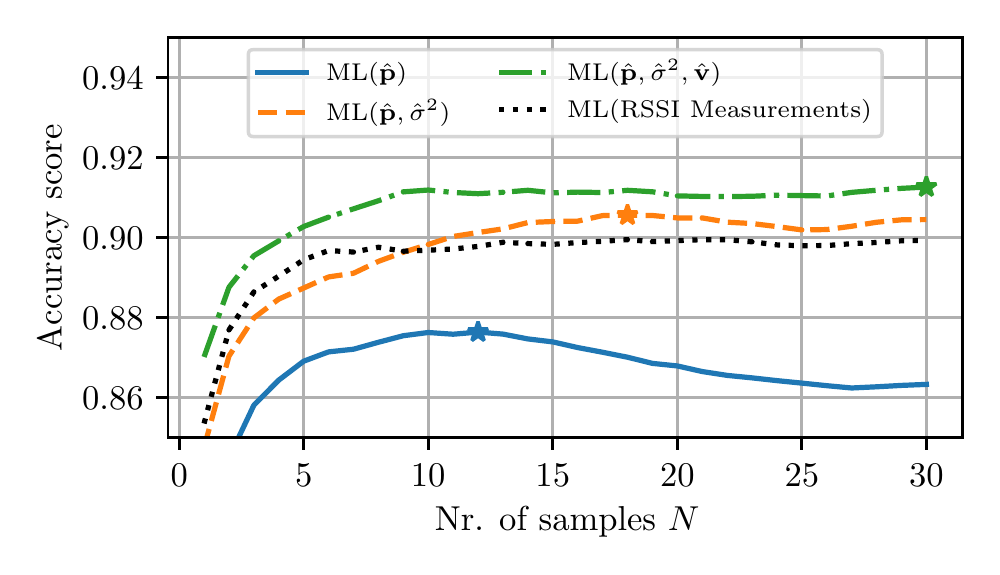}}
  \subfloat[SVM, Power Transformer]{ \includegraphics[width=.5\linewidth]{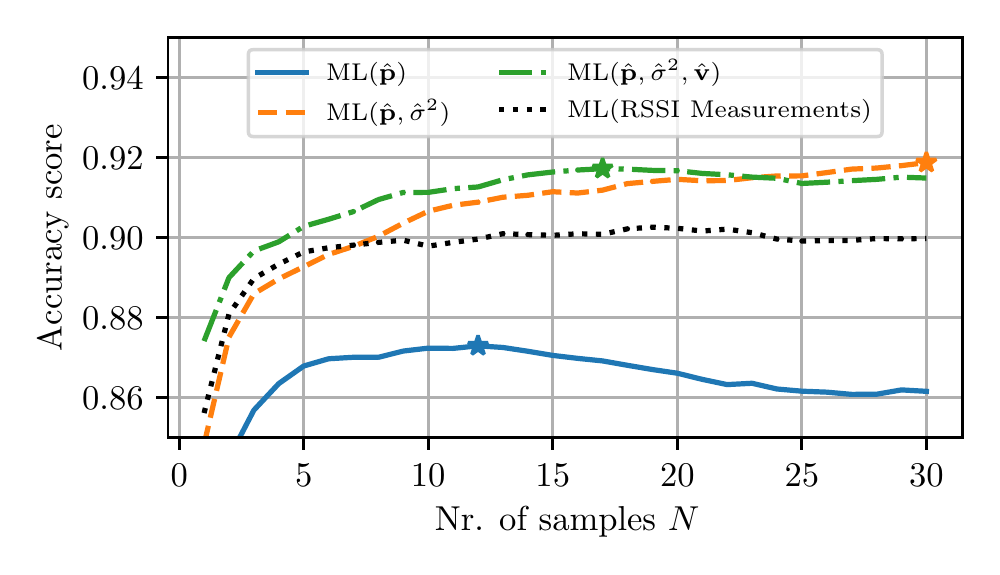}}

  \subfloat[KNN, Standard Scaler]{\includegraphics[width=.5\linewidth]{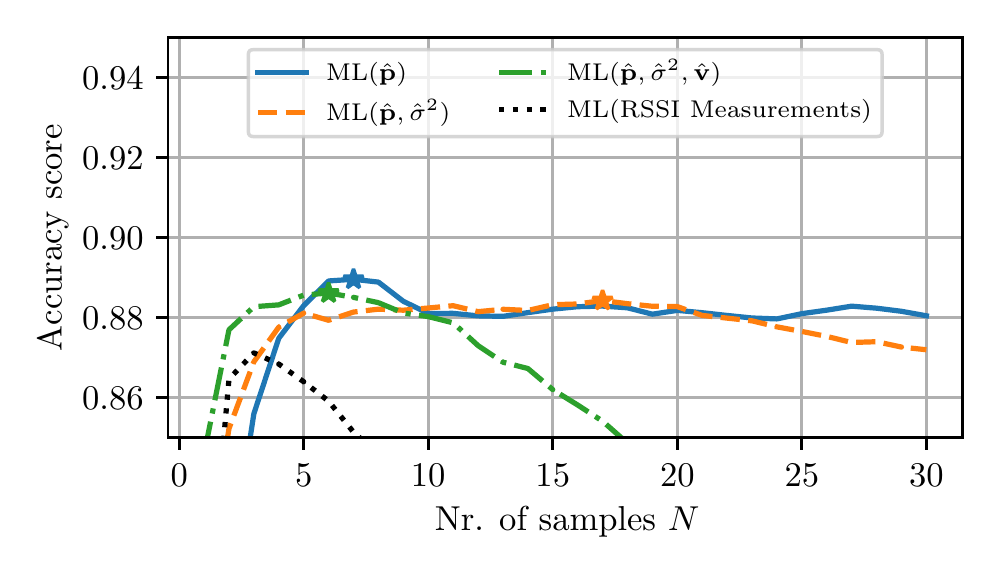}}
  \subfloat[KNN, Power Transformer]{ \includegraphics[width=.5\linewidth]{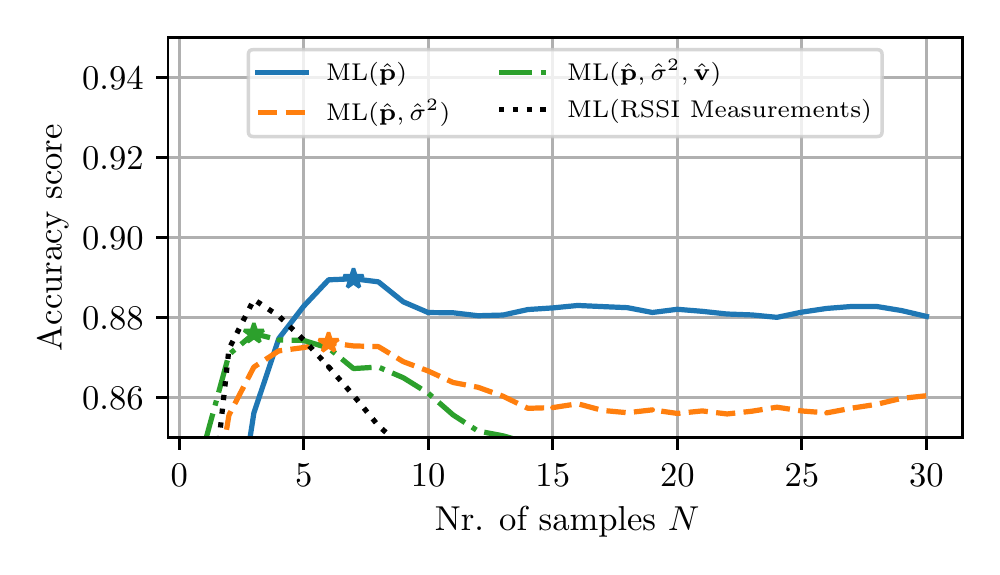}}

  \subfloat[RF, Standard Scaler]{\includegraphics[width=.5\linewidth]{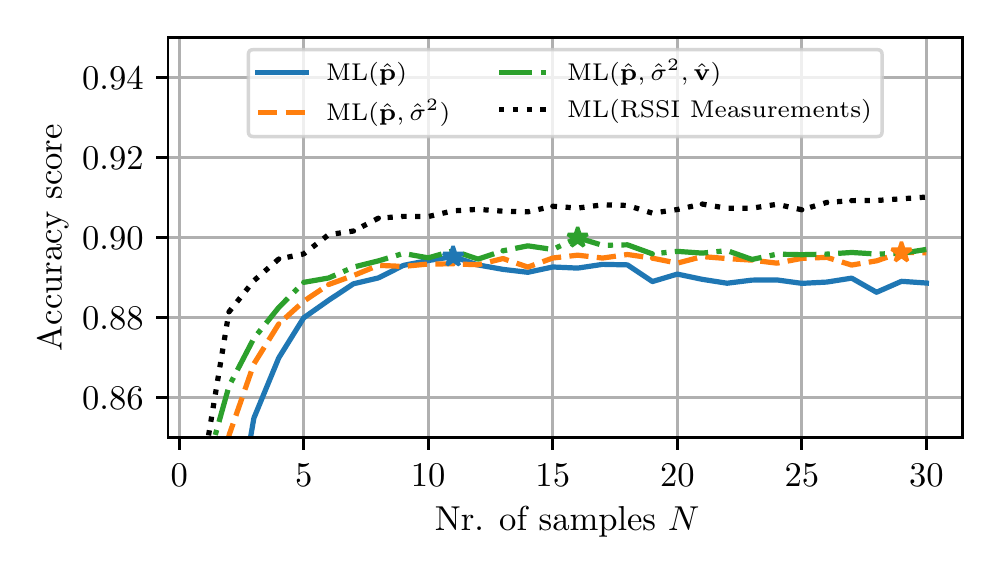}}
  \subfloat[RF, Power Transformer]{ \includegraphics[width=.5\linewidth]{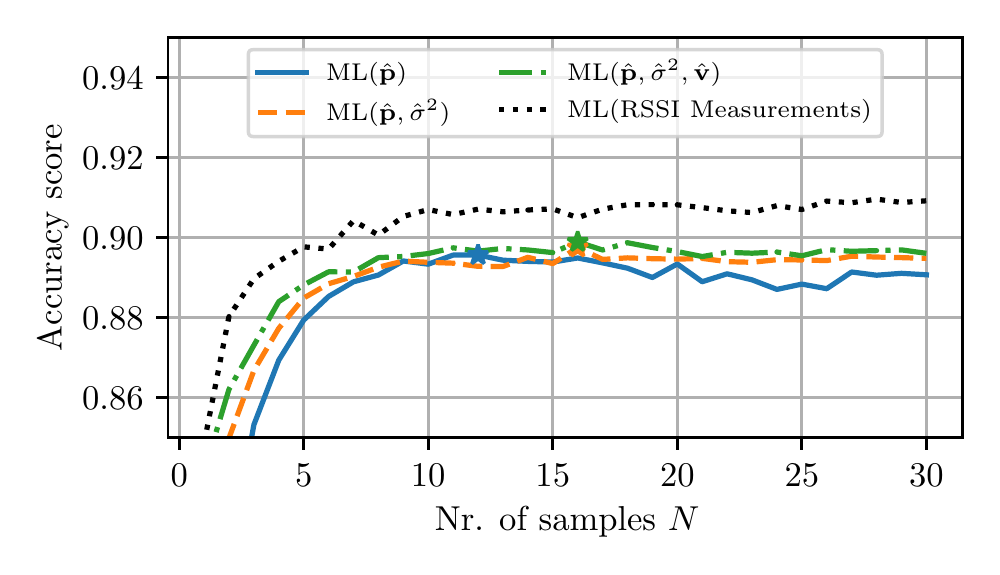}}
  \caption{Accuracy scores of the \ac{ML} classified estimates. Note the scaling of the \(y\)-axis. Locations of maxima are denoted with stars.\label{fig:svc_acc}}
\end{figure*}
The pathloss model as given in \cref{eq:pathloss} contains the term \(P_{tx} - PL_0\), which, excluding the antenna pattern, refers to the expected received power at \(1\)\,m distance. As the cars closest position to all sensors has distances of roughly \(1\)\,m, we set this term in the likelihood function to the maximum measured RSSI value of a given sensor. The pathloss exponent is set to \(n=2\), reflecting the strong line-of-sight components that we expect. The variance of the likelihood function is set to \(\sigma^2=9\).

The gain of the antenna pattern is given as
\begin{equation}
  \pi(\mtx p, \mtx q) =  \left\{
  \begin{array}{rrcl}
    0\,\text{dB},\qquad   & 0 \le              & |\alpha| & < \frac{\pi}{3}  \\
    -6\,\text{dB},\qquad  & \frac{\pi}{3} \le  & |\alpha| & < \frac{3\pi}{4} \\
    -10\,\text{dB},\qquad & \frac{3\pi}{4} \le & |\alpha| & < \pi
  \end{array}\right.,
\end{equation}
with the angle calculated as
\begin{equation}
  \alpha =\cos^{-1}\left(\frac{p^x - q^x}{\|\mtx p - \mtx q\|}\right)\,.\label{eq:directional}
\end{equation}

These assumptions are kept simple on purpose, in the hope that the sensor fusion allows to correct slight inaccuracies.
%\subsection{Results of the particle filtering}
\Cref{fig:pf_results} demonstrates the output of the location estimation for one given measurement. The ground truth is computed from the sensor information and cameras. \Cref{fig:pf_results}a shows the performance with an omnidirectional assumption, which fails to estimate the off-center positions well. \Cref{fig:pf_results}b illustrates that the directional antenna pattern is effective at correcting for these errors.
\begin{figure*}[t]
  \centering
  \subfloat[SVM, Standard Scaler, N=3]{\includegraphics[width=0.5\linewidth]{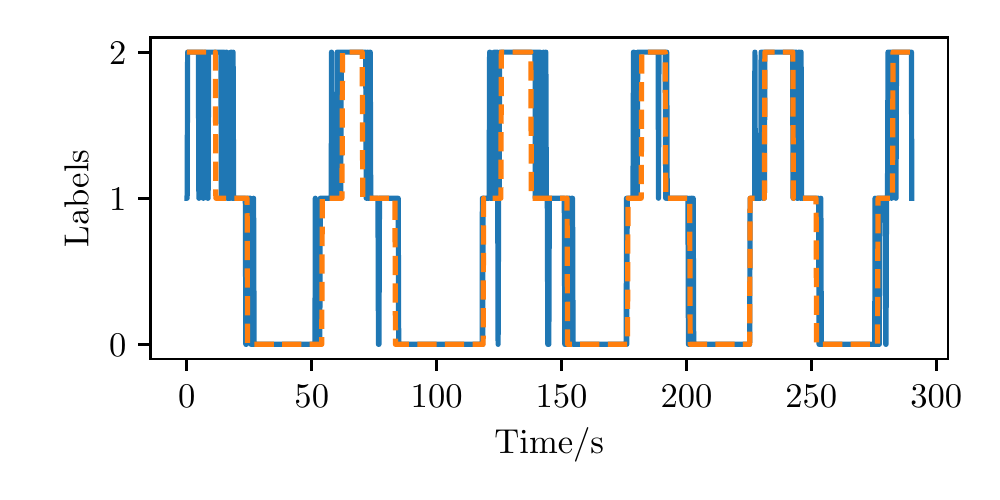}}
  \subfloat[SVM, Standard Scaler, N=16]{\includegraphics[width=0.5\linewidth]{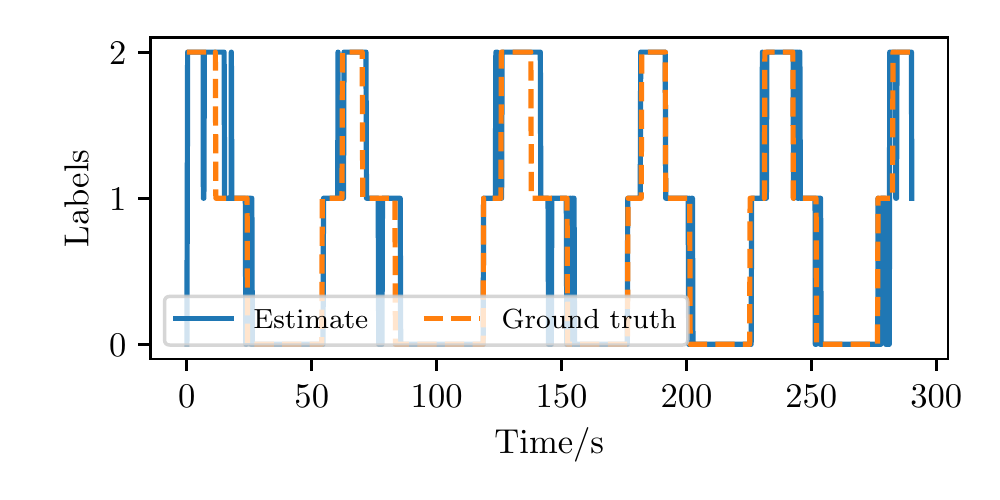}}

  \subfloat[SVM, Power Transformer, N=3]{\includegraphics[width=0.5\linewidth]{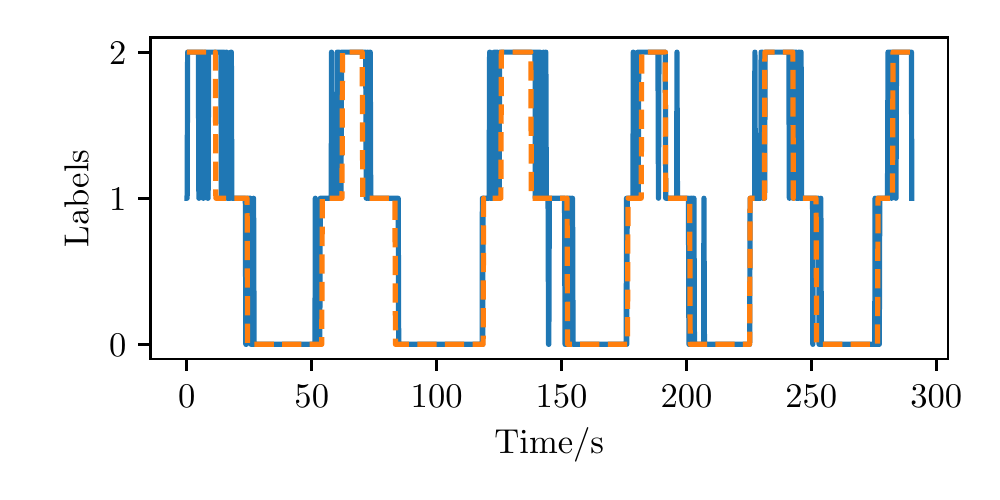}}
  \subfloat[SVM, Power Transformer, N=16]{\includegraphics[width=0.5\linewidth]{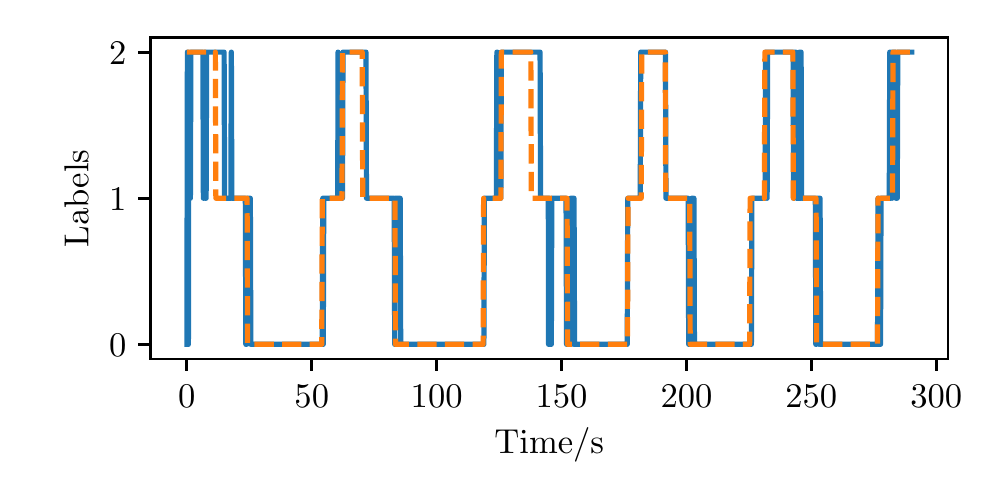}}
  \caption{Example output of SVM(\(\hat{\mtx p}, \hat{\mtx \sigma}^2, \hat{\mtx v}\)) for different memory lengths, and scalers.}\label{fig:demo-classification}
\end{figure*}
\section{ML-Based Position Detection}
\label{sec:svm}
\subsection{Machine Learning Setup}
\begin{table}[htbp]
  \caption{Classifiers with parameters\label{tab:classification_params}}
  \begin{center}
    \begin{tabular}{r|l}
      \toprule
      Estimator           & Parameters                          \\
      \midrule
      Random Forest       & n\_estimators=100, criterion="gini" \\
      K Nearest Neighbors & n\_neighbors=5                      \\
      SVM                 & kernel="rbf"                        \\
      \bottomrule
    \end{tabular}
  \end{center}

\end{table}
Based on the position and velocity estimates of the particle filter, we conduct the classification of three states as described in \cref{sec:system_model}. To this end, we employ purposefully simple \ac{ML} techniques. We consider three typical algorithms, namely \emph{\ac{KNN}}, a \emph{\ac{RF}}, and a \emph{\ac{SVM}}. The specific parameters of the \ac{ML} classifiers in the used library \emph{Scikit-learn}\cite{scikit-learn}  are given in \cref{tab:classification_params}. We combine these with three choices of data scalers, the \emph{standard scaler}, the \emph{robust scaler}, and the \emph{power transformer}. As feature vectors, we consider the position and velocity estimates \(\hat{\mtx p}\) and \(\hat{\mtx v}\) derived from \(\hat{\mtx{x}}\), as well as the variance estimate of the position estimate \(\hat{{\sigma}}^2_p\).
Additionally, to enhance the estimation, we consider a short-term history of the features. We do this by constructing a Toeplitz-matrix of width \(N\) out of each feature vector, and use the columns of the matrices as individual feature vectors.

%\subsection{Evaluation of the classification}

\begin{table}
  \begin{center}
    \caption{Accuracy Scores.\label{tab:accuracy_scores}}
    \begin{tabular}{rl|lll}
      \toprule
      Model                & Scaler            & $\mu^x$ & $ \mu^x, \sigma^x $ & $\mu^x, \sigma^x,\mu^v$ \\
      \midrule
      \multirow{3}{*}{SVM} & Standard Scaler   & 0.8764  & 0.9056              & 0.9126                  \\
                           & Robust Scaler     & 0.8764  & 0.8971              & 0.9051                  \\
                           & Power Transformer & 0.873   & 0.9188              & 0.9172                  \\
      \midrule
      \multirow{3}{*}{KNN} & Standard Scaler   & 0.8896  & 0.8843              & 0.8862                  \\
                           & Robust Scaler     & 0.8896  & 0.8586              & 0.8483                  \\
                           & Power Transformer & 0.8897  & 0.8738              & 0.876                   \\
      \midrule
      \multirow{3}{*}{RF}  & Standard Scaler   & 0.8953  & 0.8963              & 0.9                     \\
                           & Robust Scaler     & 0.8944  & 0.8967              & 0.8987                  \\
                           & Power Transformer & 0.8957  & 0.8975              & 0.8989                  \\
      \bottomrule
    \end{tabular}
  \end{center}
\end{table}

For the evaluation, we draw all possible combinations of three of the six data sets to train the \ac{ML} classifiers. Afterwards, we validate the fits against all remaining three data sets individually and compute the accuracy score defined as the fraction of correctly identified labels over the number of all instances. We don't split and shuffle the measurement runs, but instead use them as a whole either in training or testing. This approach avoids overfitting dominant measurement runs.
\subsection{Results}
\Cref{fig:svc_acc} shows the performance of the \ac{ML} classifiers for different parameter combinations, and different classifier-scaler combinations. The different curves show \ac{ML} classifiers based on either the position estimate, the position and corresponding variance estimate, or position, velocity, and variance estimates. \(N\) denotes the memory size, which is the number of samples that are considered per feature. Additionally, the plot depicts the results of the corresponding ML classifier using the raw sensor values without the particle filter preprocessing. Similarly, \cref{tab:accuracy_scores} shows the accuracy scores for each combination of input features, scaler and model if the optimum memory depth is chosen. As the table shows, the robust scaler uniformly performs worst, hence it has been ommited in \Cref{fig:svc_acc}. Furthermore, for our application, the SVM proves to perform uniformly better than the RF, while KNN performs by far the worst. Among the SVM classifiers using only the position estimate results in an overall bad behaviour. However, by adding at least the variance estimate, or even better, both variance and velocity estimate, the estimator drastically improves. The main reason for this difference is the transition region in the door. Here, the position estimate can be relatively uncertain, and will lead to many misclassifications. On the other hand, by including variance and velocity estimates, this estimate becomes much more robust. We furthermore see that in our case, the power transforming prescaler provides the overall best performance, and gains $0.5\,\%$ accuracy compared to the standard scaler.

\Cref{fig:demo-classification} illustrates the classification quality of this classifier with memory depths of three and \(16\), against the reference labels. Here, we see the discussed effect, that the transitioning label $1$ proves to be the most challenging. Both standard and power scaler show strong oscillations in the transition regions when using a memory length of $N=3$.  When increasing the memory length to $16$, it is those regions that see the most improvement.
\section{Conclusions}
\label{sec:Conclusions}
Using a particle filter as an intermediate step before machine learning allows fusing multiple low-quality feature sources into a high-quality feature source. Our results show that deriving location, and velocity mean and variance estimates and using them as features improves the performance of many \ac{ML}  classifiers, while simultaneously reducing the number of input features of the classifier. In our scenario, the best results were achieved using a SVM classifier with a power transforming prescaler.

The provided results, based on real-world measurements, prove the viability of such hybrid approaches. Additionally, doing the sensor fusion before the classification opens up the possibility of rearranging the sensor setup, and adapting the fusion process, while leaving the trained classification unchanged.
\bibliographystyle{IEEEbib}\balance
\bibliography{strings,refs,InSecTT}

\end{document}